\documentclass[twocolumn,letter]{jpsj3}
\usepackage{graphicx}
\usepackage{dcolumn}
\usepackage{bm}
\usepackage{color}
\usepackage{ulem}
\usepackage{amssymb}
\usepackage{amsmath}
\usepackage{dcolumn}
\usepackage{longtable}
\usepackage{color}
\usepackage[version = 4]{mhchem}

\newcommand{\lrp}{LaRu$_4$P$_{12}$}

\newcommand{\pnmr}{$^{31}$P}
\newcommand{\lnmr}{$^{139}$La}

\begin{document}
\title{Magnetic field dependence of the spin susceptibility on conventional $s$-wave superconductor \lrp\ revealed by \pnmr-NMR and \lnmr-NMR}
\author{Riku~Matsubayashi$^{1,}$\thanks{E-mail address: matsubayashi.riku.38a@st.kyoto-u.ac.jp},
Shiki~Ogata$^{1}$,
Taishi~Ihara$^{1}$,
Hiroyasu~Matsudaira$^{1}$,
Shunsaku~Kitagawa$^{1}$,
Kenji~Ishida$^{1}$,
Yusuke~Nakai$^{2}$,
Hitoshi~Sugawara$^{3}$,
Hideyuki~Sato$^{4}$
}
\inst{$^1$Department of Physics, Graduate School of Science, Kyoto University, Kyoto 606-8502, Japan \\
$^2$Graduate School of Material Science, University of Hyogo, Hyogo 678-1297, Japan \\
$^3$Department of Physics, Graduate School of Science, Kobe University, Hyogo 657-8501, Japan \\
$^4$Department of Physics, Tokyo Metropolitan University, Hachioji, Tokyo 192-0397, Japan \\
}
\date{\today}
\abst{
The magnetic field dependence of the spin part of Knight shift, which is proportional to the superconducting-state spin susceptibility, was investigated at two nuclear sites, \pnmr\ and \lnmr\ in a conventional $s$-wave superconductor \lrp. 
After the analyses, we confirmed that the superconducting-state spin susceptibility is proportional to magnetic field, and connects to the normal-state spin susceptibility smoothly. This is a textbook example, when the superconductivity is broken with the orbital pair-breaking effect.
}

\maketitle
In superconductors, two electrons in the crystal form Cooper pairs, which can be either spin-singlet state with antiparallel spins\cite{PhysRev.108.1175,PhysRev.110.769,Fine1969,S.Kitagawa_PRB_2017,S.Kitagawa_PRL_2023} or spin-triplet state with parallel spins\cite{PhysRevLett.77.1374,PhysRevLett.80.3129,doi:10.7566/JPSJ.88.064706,A.J.Leggett_RevModPhys_1975,H.Matsumura_JPSJ_2023,J.Yang_SciAdv_2021}. 
Most superconductors, including \lrp, are in the spin-singlet pairing. 
In the spin-singlet state, there are two main mechanisms contributing to the break of Cooper pairs under magnetic field: the Pauli pair-breaking effect\cite{Pauli1,Pauli2,K.Maki_PTP_1964} and the orbital pair-breaking effect\cite{WHH}. 
In the Pauli pair-breaking effect, the Zeeman splitting energy due to the external magnetic field exceeds the condensation energy of the superconducting (SC) state, thereby destroying the Cooper pairs\cite{T.Tayama_PRB_2002,Y.Maeno_JPSJ_2024}. 
On the other hand, in the orbital pair-breaking effect, the Lorentz force resulting from the applied external magnetic field exceeds the binding force between two electrons and destroys the pairs\cite{S.Uji_JPSJ_2023,M.Hagawa_JPSJ_2024}.

Investigating the magnetic field $H$ dependence of the SC spin susceptibility gives various insights into the SC symmetry. 
For instance, when the Pauli pair-breaking effect is dominant, a discontinuous change in the $H$ dependence of the spin susceptibility or the formation of spatially oscillating Fulde‑Ferrell‑Larkin‑Ovchinnikov state has been observed near the upper critical field $\mu_{0}H_{\text{c2}}$ in various strongly correlated electron superconductors\cite{B.L.Young_PRL_2007,S.Kitagawa_JPSJ_2017,S.Kitagawa_PRL_2018,doi:10.1073/pnas.2025313118,K.Kinjo_science_2022}. 
In contrast, when a spin-triplet state is realized, it has been unveiled that the decrease in the spin susceptibility is suppressed under $H$ smaller than $H_{\text{c2}}$, from recent experiments on UPt$_3$\cite{PhysRevLett.77.1374,PhysRevLett.80.3129,doi:10.7566/JPSJ.88.064706},UTe$_2$\cite{G.Nakamine_JPSJ_2021,G.Nakamine_PRB_2021,K.Kinjo_PRB_2023,S.Kitagawa_JPSJ_2024} and K$_2$Cr$_3$As$_3$\cite{J.Yang_SciAdv_2021}. 
On the other hand, precise measurements of the spin susceptibility against $H$ in the conventional superconductors have rarely been conducted. 
This is because the spin susceptibility in the SC state is difficult to evaluate, as discussed later. 
Theoretically, the quasiparticle density of states around Fermi energy $D(E_{\text{F}})$ in the mixed state is proportional to $H/H_{\text{c2}}$. 
In fact, a linear change in the density of states was reported from $H$ dependence of the specific heat and scanning tunneling spectroscopy measurements\cite{M.Nohara_JPSJ_1999,HANAGURI2003}. 
As the spin susceptibility $\chi_{\text{spin}}$ is proportional to $D(E_{\text{F}})$, it is expected $\chi_{\text{spin}}$ in the SC state would be expressed as a linear function of $H$. 

\lrp $ $ is known to be an $s$-wave superconductor with an isotropic SC energy gap from previous studies\cite{G.P.Meisner_Physica_1981,I.Shirotani_JPCS_1996,S.Tsuda_JPSJ_2006}, exhibiting a Hebel-Slichter peak just below the SC transition temperature 7.2 K\cite{Y.Nakai_PhysicaB_2008,Y.Nakai_JPSJ_sa_2008,K.Kinjo_JPSJ_2019}. 
From $H$ dependence of $\mu_{0}H_{\text{c2}}$, as shown in Fig.~1 (c), \lrp $ $ is considered to be a typical example of a superconductor where $H_{\text{c2}}$ is determined by the orbital pair-breaking effect\cite{I.Shirotani_JPCS_1996,K.Kinjo_JPSJ_2019}, since the $dH_{\rm c2}/dT$ is linear almost up to $T \sim 0 \ \text{K}$, and the strong suppression of $H_{\rm c2}$, the characteristics of the Pauli pair-breaking effect, was not observed at all, as seen in Fig.~1 (c). 
In addition, the dominance of the orbital pair-breaking effect is also suggested from the measurement of the nuclear spin-lattice relaxation rate $1/T_{1}T$. Since $1/T_{1}T$ is proportional to $D(E_{\text{F}})^{2}$, the relation of $1/T_{1}T \propto H^{2}$ in the SC state below $\mu_{0}H_{\text{c2}}$\cite{K.Kinjo_JPSJ_2019} was revealed in \lrp. 
Since the spin susceptibility $\chi_{\text{spin}}$ in the SC state is proportional to $D(E_{\text{F}})$, $\chi_{\text{spin}}$ would be proportional to $H$ in \lrp.
In fact, when $H$ dependence of the spin susceptibility is measured by NMR measurements, the influence of the SC diamagnetism becomes dominant at low $H$, making it difficult to derive the contribution of the spin susceptibility alone. 
Particularly, the spectrum broadening by electric quadrupole interaction and the SC diamagnetism make the experimental resolution worse and more complicated. 

\lrp $ $ has a body-centered cubic structure and a cubic symmetry $Im\overline{3}$ space group\cite{W.Jeitschko_AC_1977}, as shown in Fig.~1 (a). 
In addition, NMR measurements can be performed, and a sharp NMR peak is observed at both the \pnmr $ $ and \lnmr $ $ sites. 
Therefore, from the precise Knight shift measurements at both sites, it is possible to evaluate $H$ dependence of the spin susceptibility irrespective of the SC diamagnetism.

In this paper, we performed \pnmr $ $ and \lnmr -NMR measurements and subtracted SC diamagnetism experimentally, as carried out previously\cite{PhysRevLett.74.1649,doi:10.1143/JPSJ.74.3370}. 
We revealed that the SC spin susceptibility above 1 T showed a $H$-linear dependence and smoothly connected to the normal-state spin susceptibility $\chi_{\text{normal}}$. These are quite consistent with the theoretical expectation for the orbital pair-breaking scenario.

\begin{figure}[t]
    \begin{center}
    \includegraphics[]{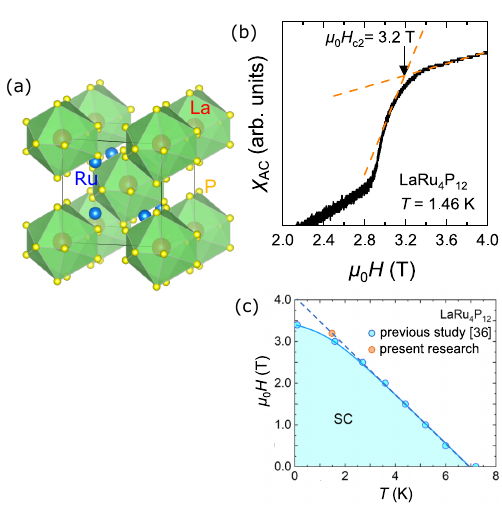}
    \caption{(Color online) (a) Crystal structure of \lrp\ drawn by VESTA\cite{K.Momma_JAC_2011}. (b) Magnetic field dependence of $\chi_{\text{AC}}$ at 1.46 K. The arrow indicates $\mu_{0}H_{\text{c2}}=3.2$ T. (c) $H$--$T$ phase diagram of \lrp $ $. The blue points are from the previous study\cite{K.Kinjo_JPSJ_2019} and the orange point is from the present measurement.}
    \end{center}
\end{figure}

The sample \lrp $ $ used in the present experiment was synthesized by the Sn-flux method\cite{T.Uchiumi_JPCS_1999}. 
The sample was powdered to enhance the NMR-signal intensity and enclosed in a straw case, the same as used in the previous study\cite{K.Kinjo_JPSJ_2019}. 
The NMR measurements were conducted at the \pnmr $ $ site (the nuclear spin $I =1/2$, gyromagnetic ratio $^{31} \gamma /2 \pi=17.235$ MHz/T) and the \lnmr $ $ site (the nuclear spin $I =7/2$, gyromagnetic ratio $^{139} \gamma /2 \pi=6.0142$ MHz/T). 
The standard spin-echo technique was used. The NMR spectra were obtained by performing a Fourier transform on the spin-echo signals following radio-frequency (RF) pulses under multiple fixed magnetic fields ranging from 0.5 T to 4.0 T. 
Magnetic field calibration was carried out using $^{63}$Cu ($^{63} \gamma /2 \pi=11.289$ MHz/T) and $^{65}$Cu ($^{65} \gamma /2 \pi=12.093$ MHz/T) signal arising from the NMR coil for several magnetic fields. 
Here we assumed that normal-state Knight shift is independent of $H$, which is usual in conventional metals.
The NMR spectra in the SC state were observed using a field-cooling process to avoid the random vortex distribution. 
$T_{\text{c}}$ and the upper critical field $\mu_{0} H_{\text{c2}}$ were determined from the ac measurements of magnetic susceptibility $\chi_{\text{AC}}$ using the NMR tank circuit as shown in Fig.~1 (b). In the SC state, the impedance of the circuit changes due to the Meissner effect and thus the tuning frequency of the circuit significantly changes just below $\mu_{0} H_{\text{c2}}$. 
The upper critical field $\mu_{0} H_{\text{c2}}$ was 3.2 T at $T$=1.46 K from the measurement of $\chi_{\text{AC}}$.

The Knight shift values were determined at the peak position in each NMR spectrum. The Knight shift at 10 K was used for the normal-state value, while the Knight shift at 1.4 K was adopted for the superconducting-state value. The error bar for the Knight shift value was taken as the width at 99 \% of the peak intensity of the spectrum. 

\begin{figure}[t]
    \begin{center}
    \includegraphics[]{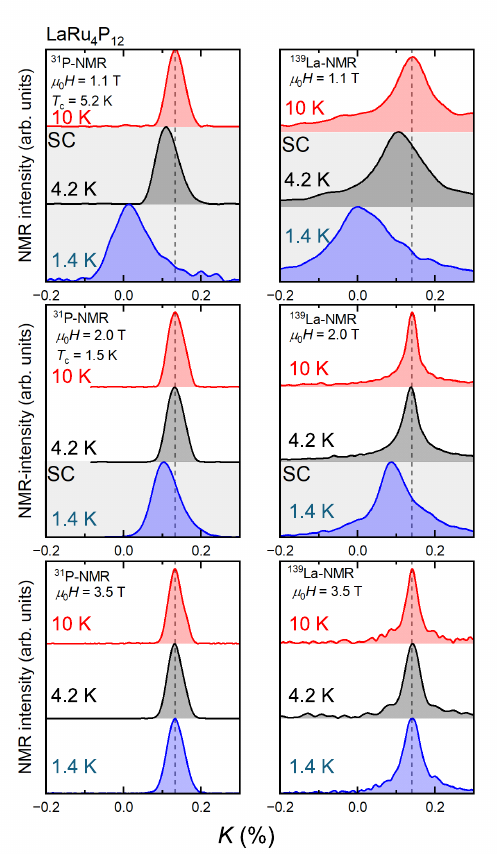}
    \caption{(Color online) Knight shift spectra at $\mu_{\text{0}}H = 1.1$ T, 2.0 T and 3.5 T. Measurements were performed at three temperatures ranging from 10 K to 1.4 K. The dashed lines represent the peak Knight shift at 10 K. The background color of spectra in the SC state is gray. At lower temperatures, the peak Knight shift of the spectrum decreases, and at the lowest temperature, the spectrum becomes broadened.}
    \end{center}
\end{figure}

Figures 2 shows the typical \pnmr-$ $ and \lnmr-NMR spectra measured under the applied magnetic field of $\mu_{0}H=1.1$ T, 2.0 T, and 3.5 T. 
The horizontal axis is converted to $K=(f-f_{0})/{f_{0}}$. 
Here $f_{0}$ is the reference frequency at $K$ = 0, which is determined by $f_{0}=\gamma_n H/{2 \pi}$ in each nucleus.
The measurements were conducted at three temperatures: 10 K, 4.2 K, and 1.4~ K, at both sites.
Below $T_{\text{c}}$($H$), the peak positions of the spectra decrease. 
The shapes of the spectra are sharper and symmetric at temperatures above $T_{\text{c}}$ for both the \pnmr $ $ and \lnmr $ $ sites. Here, we observed single sharp \lnmr -NMR spectra without a quadrupolar eﬀect, as observed in LaFe$_{4}$P$_{12}$\cite{doi:10.1143/JPSJ.74.3370}, since the \lnmr $ $ nucleus (nuclear spin $I$ = 7/2) is located at the cubic-symmetry site.
At 1.1 T, the spectra become asymmetric and broadened with tails towards the higher $K$ side at 1.4 K. 
This shape is consistent with the NMR spectra reported in the previous results in the SC state\cite{Y.Nakai_JPSJ_sa_2008,Y.Nakai_PhysicaB_2008}.
The asymmetry in the spectra is attributed to the so called "Redfield pattern", a characteristic NMR spectrum in the SC vortex state\cite{Redfield_PRL_1966,Y.Nakai_JPSJ_sa_2008}. 
This Redfield pattern arises from the distribution of the internal field and Knight shift in the SC vortex lattice.
Moreover, the negative Knight shift in the spectrum with the Redfield pattern at 1.4 K is mainly due to the effect of the SC diamagnetism. 
On the other hand, at 2.0 T, the asymmetry of the spectrum becomes smaller. 
When $H$ is larger than $H_{\text{c2}}$, the spectrum does not change at all down to 1.4 K. 
$H$ dependence of $K$ at 1.4 K and 10 K is plotted in Figs. 3 (a) and 3 (b) for the \pnmr $ $ site and the \lnmr $ $ site, respectively. 
The Knight shift of \pnmr $ $ and \lnmr-NMR at 1.4 K increases with $H$, while $K$ at 10 K is almost constant against $H$.
\begin{figure}[t]
    \begin{center}
    \includegraphics[scale=0.79]{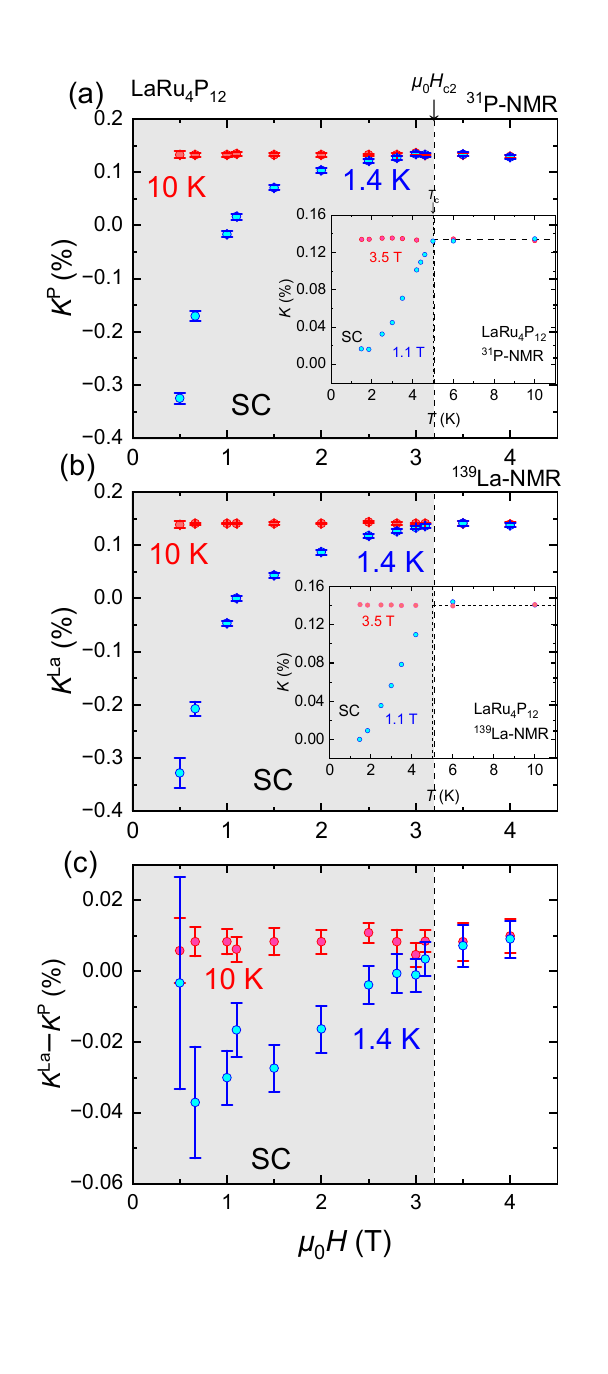}
    \caption{(Color online) $H$ dependence of $K$ at 1.4 K and 10 K for \pnmr $ $  (a) and \lnmr $ $ site (b). (Inset) $T$ dependence of $K$ at 1.1 T and 3.5 T. (c) $H$ dependence of $\Delta K(=K^{\text{La}}-K^{\text{P}})$ at 10 K and 1.4 K. }
    \end{center}
    \end{figure}

In the SC state, the measured Knight shift $K(T, H)$ can be expressed as
\begin{gather}
K(T,H) = K_{\text{spin}}(T,H) + K_{\text{orb}} + K_{\text{dia}}(T,H).
\end{gather}
Here, $K_{\text{spin}}$ is the electron-spin component of the Knight shift. 
$K_{\text{orb}}$ is driven by the Van Vleck susceptibility induced by the orbital angular momentum of the electron, which is $T$ independent at low $T$. 
$K_{\text{dia}}$ arises from the diamagnetism in the SC state, and it takes a negative value and is $T$ and $H$ dependent. \\
The difference between $K(10\ \text{K})$ and $K(1.4\ \text{K})$ is expressed as

\begin{equation}
\begin{split}
\Delta K^{\text{}} &= K^{\text{}}(10\ \text{K}) - K^{\text{}}(1.4\ \text{K}) \\
&= \Delta K_{\text{spin}}^{\text{}} - K_{\text{dia}}.
\end{split}
\end{equation}

$\Delta K^{\text{}}_{\text{spin}}$ is expressed as

\begin{equation}
\begin{split}
\Delta K^{\text{}}_{\text{spin}}=K^{\text{}}_{\text{spin}}(10\ \text{K})-K^{\text{}}_{\text{spin}}(1.4\ \text{K}).
\end{split}
\end{equation}  

At lower fields, $K_{\text{dia}}$ becomes larger and comparable to $K_{\text{spin}}$, making it difficult to derive $H$ dependence of $K_{\text{spin}}$. 
Since $K_{\text{dia}}$ is a bulk effect, it is reasonable to assume that $K_{\text{dia}}$ works at the \pnmr $ $ and \lnmr $ $ sites in the same manner. 
Thus, we can subtract the contribution of $K_{\text{dia}}$ by measuring $K$ at two crystalographic different sites. 
The difference in $\Delta K$ between \lnmr $ $ and \pnmr $ $ sites is given by

\begin{equation}
\begin{split}
\Delta K^{\text{La}}-\Delta K^{\text{P}}
&= \Delta K_{\text{spin}}^{\text{La}}- \Delta K_{\text{spin}}^{\text{P}}\\
&=(A_{\text{hf}}^{\text{La}} - A_{\text{hf}}^{\text{P}})\Delta \chi_{\text{spin}} .
\end{split}
\end{equation}
 
$\Delta \chi_{\text{spin}}$ is expressed as

\begin{equation}
\begin{split}
\Delta \chi_{\text{spin}}=\chi_{\text{spin}}(10\ \text{K})-\chi_{\text{spin}}(1.4\ \text{K}).
\end{split}
\end{equation}  

Thus, we can get $\Delta K_{\text{spin}}^{i}$ ($i$ = \pnmr $ $ and \lnmr) as

\begin{equation}
\begin{split}
\Delta K_{\text{spin}}^{i} &= A^{i}_{\text{hf}}\Delta \chi_{\text{spin}} \\
&= \frac{A^{i}_{\text{hf}}}{A^{\text{La}}_{\text{hf}}-A^{\text{P}}_{\text{hf}}}(\Delta K^{\text{La}}-\Delta K^{\text{P}}).
\end{split}
\end{equation}

Here, $A^{i}_{\text{hf}}$ is the hyperfine coupling constant at the $i$ site and $\chi_{\text{spin}}$ is the spin susceptibility in \lrp. The relation of $K$ at \pnmr $ $ site and \lnmr $ $ site at each temperature (from 10 K to 220 K) is shown in Fig.~4. The dashed line represents the linear function fitted from 10 K to 220 K. Here, the slope of the linear line gives the value of  $A^{\text{P}}_{\text{hf}}/A^{\text{La}}_{\text{hf}}$. The obtained value of $A^{\text{P}}_{\text{hf}}/A^{\text{La}}_{\text{hf}}$ is 0.60. Therefore, we can evaluate $H$ dependence of $\Delta K_{\text{spin}}^{i}$ by using this value and the experimental data. 

\begin{figure}[t]
    \begin{center}
    \includegraphics[scale=0.9]{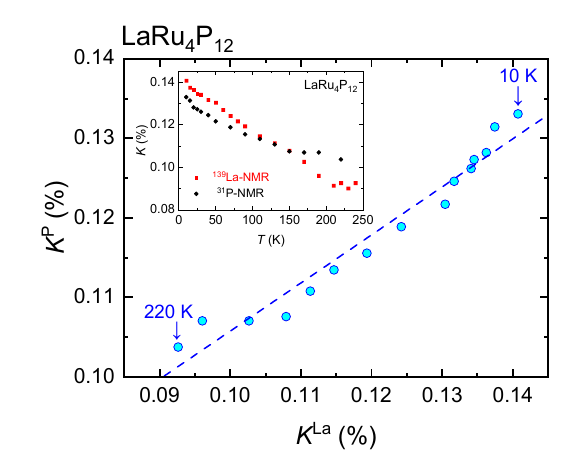}
    \caption{(Color online) The plot of $K^{\text{P}}$ against $K^{\text{La}}$ above 10 K. (Inset) $T$ dependence of $K$ at \pnmr $ $ site and \lnmr $ $ site.}
    \end{center}
    \end{figure}

\begin{figure}[t]
\begin{center}
\includegraphics[scale=0.9]{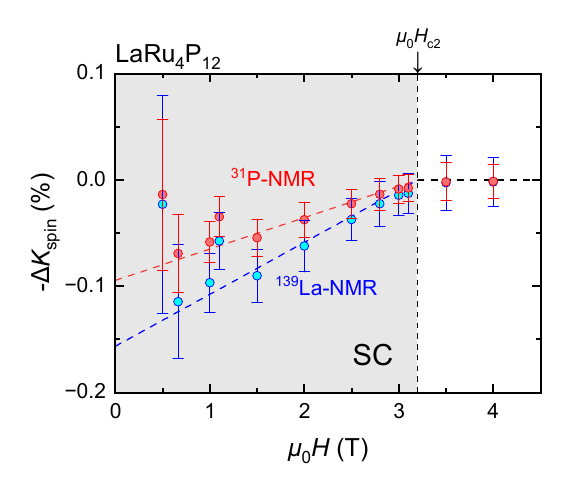}
\caption{(Color online) $H$ dependence of $\Delta K_{\text{spin}}$ at\pnmr $ $ site and \lnmr $ $ site. The background color in the in the SC state is gray. The dashed line below $\mu_{0}H_{\text{c2}}$ represents the fitting function (fitted with a linear function). The arrow indicates $\mu_{0}H_{\text{c2}}$.}
\end{center}
\end{figure}

Figure 5 shows $H$ dependence of $\Delta K_{\text{spin}}^{i}$ ($i$ = \pnmr $ $ and \lnmr). 
Up to $\mu_{0}H_{\text{c2}}$, the value increases linearly with increasing $H$, reaching the value in the normal state at $\mu_{0}H_{\text{c2}}$, which is consistent with previous result\cite{K.Kinjo_JPSJ_2019} and expectation. This implies that $\chi_{\text{spin}}$ in the SC state, which is proportional to $K_{\text{spin}}$, also linearly recovers with increasing magnetic fields.
However, the measurement at low fields ($\mu_{0}H$ = 0.5 T, 1.1 T) deviates from this behavior with large error bar.  
There is an appreciable increase in error at low fields due to the dominant contribution of the SC diamagnetism. If one extrapolates $\Delta K^{\text{}}_{\text{spin}}$ to zero magnetic field, the resulting value is nearly equal to $K^{\text{}}$(10 K). This means that the decrease in $K$ below $T_{\text{c}}$ is predominantly due to the reduction of $K_{\text{spin}}$ in the SC state.

$K_{\text{dia}}$ is also estimated from eq. (2) by using the values of $\Delta K_{\text{spin}}$. By subtracting $\Delta K_{\text{}}$ from $\Delta K_{\text{spin}}$, the value of $K_{\text{dia}}$ is obtained, as shown in Fig.~6. Here, some values of $K_{\text{dia}}$ above 2.0 T are positive because of the experimental error.

$H$ dependence of $K_{\text{dia}}$ is also approximately calculated by using the following theoretical expression\cite{E.Brandt_PRB_2003}: 

\begin{figure}[t]
    \begin{center}
    \includegraphics[scale=0.9]{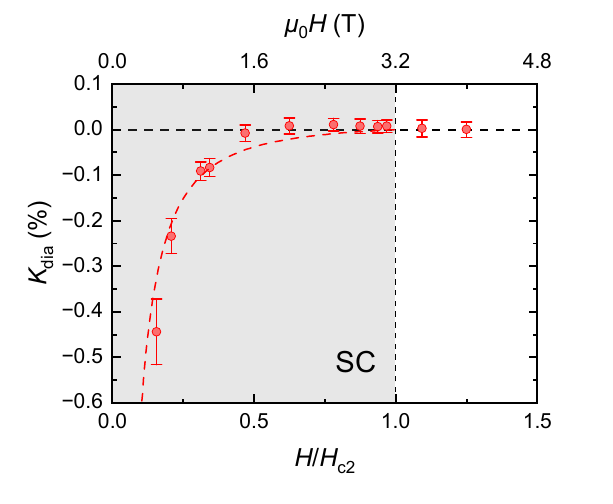}
    \caption{(Color online) $H$ dependence of estimated $K_{\text{dia}}$. The value of $\mu_{0}H$ is scaled by $\mu_{0}H_{\text{c2}}=3.2$ T. The dashed line is the fitting line of $K_{\text{dia}}$. Some values of $K_{\text{dia}}$ are positive due to the experimental error.}
    \end{center}
    \end{figure}
    
\begin{gather}
K_{\text{dia}} = -(1-N)\frac{H_{\text{c2}}}{H} \frac{\ln \left(\frac{H_{\text{c2}}}{H}\right)}{4 \kappa^{2}} \times 100 \ (\%),
\end{gather}
where $\mu_{0}H_{\text{c2}}$ is the upper critical field, $N$ (= 1/3) is the demagnetization factor and $\kappa$ is the Ginzburg-Landau parameter. By fitting the experimental data using this equation, we can get $\kappa = 24.7$ from \pnmr-NMR and \lnmr-NMR. Using $\kappa = 24.7$ and $\mu_{0}H_{\text{c2}}=3.2$ T, the value of $\mu_{0}H_{\text{c1}}$ is estimated to be 10 mT, where $H_{\text{c1}} = (H_{\text{c2}}\ln{\kappa})/{2 \kappa^{2}}$. These values are consistent with $\kappa=22$ with superconductor LaFe$_{4}$P$_{12}$\cite{doi:10.1143/JPSJ.74.3370}.

In conclusion, we performed \pnmr\ and \lnmr\ -NMR measurements on \lrp\ to investigate the magnetic field response of the spin susceptibility.
The Knight shift obtained from the \pnmr\ and \lnmr-NMR\ measurements decreases at low temperatures.
In addition, the NMR spectrum becomes asymmetric and broadened at lower temperatures. 
The spectrum change in the SC state is attributed to the Redfield pattern, which is a characteristic NMR spectrum in the vortex state. 
From NMR Knight shift, it is confirmed that the spin susceptibility $\chi_{\text{spin}}$ can be expressed with a coefficient $\alpha$ as $\chi_{\text{spin}} = \alpha (H - H_{\text{c2}})+\chi_{\text{normal}}$, which indicates a linear relationship with the magnetic field. 
This result is consistent with the previous result\cite{K.Kinjo_JPSJ_2019}, and theoretical prediction when $H_{\rm c2}$ is determined with the orbital pair-breaking effect. The relationship is actually observed in FeSe recently\cite{PhysRevB.105.054514}.

\section*{acknowledgments}
The authors would like to thank H. Matsumura, and M. Shibata for experimental support. 
This work was supported by Grants-in-Aid for Scientific Research (KAKENHI Grant No. JP20KK0061, No. JP20H00130, No. JP21K18600, No. JP22H04933, No. JP22H01168, No. JP23H01124, No. JP23K19022, No. JP23K22439 and No. JP23K25821) from the Japan Society for the Promotion of Science, by JST SPRING(Grant No. JPMJSP2110) from Japan Science and Technology Agency, by research support funding from The Kyoto University Foundation, by ISHIZUE 2024 of Kyoto University Research Development Program, by Murata Science and Education Foundation, and by the JGC-S Scholarship Foundation.
In addition, liquid helium is supplied by the Low Temperature and Materials Sciences Division, Agency for Health, Safety and Environment, Kyoto University.

\bibliographystyle{jpsj_QM}  
\bibliography{LaRu4P12}

\end{document}